\def\la{${_<\atop^{\sim}}$}
\def\ga{${_>\atop^{\sim}}$}
\def\ad{$''\!{.}$} 
\def\dd{$^{\circ}\!{.}$} 
\def\deg{$^{\circ}$} 
\def\msun{$M_{\odot}$} 
\documentstyle[12pt]{article}

\begin{document}
\begin{typestyle}
\normalsize
\begin{center}
{\bf THE VERY LOW MASS COMPONENT OF THE GLIESE 105 SYSTEM}\\[0.75in]
D{\small AVID}~A.~G{\small OLIMOWSKI}$^1$, 
T{\small ODD}~J.~H{\small ENRY}$^1$,
J{\small OHN}~E.~K{\small RIST}$^2$,
D{\small ANIEL}~J.~S{\small CHROEDER}$^3$,
G{\small EOFFREY}~W.~M{\small ARCY}$^4$, 
D{\small EBRA}~A.~F{\small ISCHER}$^4$, \\
{\small AND} 
R.~P{\small AUL}~B{\small UTLER}$^5$
\\[1in]
\end{center}
\noindent
$^1${\small Department of Physics and Astronomy, The Johns Hopkins University, Baltimore, MD 21218} \\[0.10in]
$^2${\small Space Telescope Science Institute, 3700 San Martin Drive, Baltimore, MD 21218} \\[0.10in]
$^3${\small Department of Physics and Astronomy, Beloit College, Beloit, WI 53511} \\[0.10in]
$^4${\small Department of Astronomy, University of California, Berkeley, CA 94720 and Department of Physics and Astronomy, San 
Francisco State University, San Francisco, CA 94132}\\[0.1in]
$^5${\small Department of Terrestrial Magnetism, Carnegie Institution of Washington, 5241 Broad Branch Road NW, Washington, DC 
20015} \\[1.00in]
\hspace*{0.5in} To be published in the October 2000 issue of {\it The Astronomical Journal}. \\[0.5in]
\hspace*{0.5in} Submitted : \hspace*{0.2in} 2 May 2000 \\[0.2in]
\hspace*{0.5in} Accepted :  \hspace*{0.3in} 16 June 2000 \\[0.5in]
\pagebreak
\vspace*{0.25in}
\begin{center}
{\bf ABSTRACT} \\[0.2in]
\end{center}

\noindent
Multiple-epoch, multicolor images of the astrometric binary Gliese~105A and its very low mass companion Gliese~105C have been 
obtained using the {\it Hubble Space Telescope's} Wide Field Planetary Camera 2 (WFPC2) and Near-Infrared Camera and Multi-Object
Spectrometer (NICMOS).  The optical and near-infrared colors of Gl~105C strongly suggest a spectral type of M7~{\small V} for 
that star.  Relative astrometric measurements spanning 3~yr reveal the first evidence of Gl~105C's orbital motion.  Previous 
long-term astrometric studies at Sproul and McCormick Observatories have shown that the period of Gl~105A's perturbation is 
$\sim 60$~yr.  To satisfy both the observed orbital motion and Gl~105A's astrometric period, Gl~105C's orbit must have an 
eccentricity of $\sim 0.75$ and a semimajor axis of $\sim 15$~AU.  Measurements of Gl~105A's radial velocity over 12~yr show a 
linear trend with a slope of 11.3~m~s$^{-1}$~yr$^{-1}$, which is consistent with these orbital constraints and a nearly face-on 
orbit.  As no other faint companions to Gl~105A have been detected, we conclude that Gl~105C is probably the source of the 60-yr 
astrometric perturbation.\\


\pagebreak 
\begin{center}
{\bf 1. INTRODUCTION} \\
\end{center}

Gliese 105 is a visual triple system comprising a K3~{\small V} primary star (Gl~105A, HR~753, HD~16160, BD~+6\deg398; $V = 
5.82$), a M3.5~{\small V} secondary star (Gl~105B; $V = 11.7$) located $165''$ to the southeast (van~Maanen 1938), and a very low
mass (VLM) tertiary star (Gl~105C; $V = 16.8$) located $\sim 3$\ad3 to the northwest of Gl~105A (Golimowski et al.\ 1995b, 
hereafter GNKO; Golimowski et al.\ 1995a, hereafter GFSU).  Long-term astrometric studies at Sproul Observatory (Lippincott 1973;
Heintz \& Cantor 1994) and McCormick Observatory (Martin \& Ianna 1975; Ianna 1992) indicate that Gl~105A suffers a perturbation 
with a period of $\sim 60$~yr.  Radial-velocity measurements of Gl~105A over the last 12~yr reveal a long-term trend with a 
linear slope of $11.3 \pm 0.8$~m~s$^{-1}$~yr$^{-1}$ (Cumming, Marcy, \& Butler 1999; this paper).  The position of Gl~105C 
observed in 1995 differs greatly from the positions predicted for that epoch from the latest published orbital elements of the 
astrometric companion (GNKO, GFSU).  The discrepancies between the observed and predicted positions have fueled speculation that 
the perturbation of Gl~105A may be caused by an unseen fourth component to the Gl~105 system.  However, direct images of Gl~105A 
obtained with the {\it Hubble Space Telescope (HST)} reveal no other companions as faint as the coolest known brown dwarfs lying 
within $10''$--$17''$ of the star (Schroeder et al.\ 2000; this paper).\\

The small separation and large brightness ratio of Gl~105AC render photometry and spectroscopy of Gl~105C difficult.  GFSU found
that the $V$--$I$ color of Gl~105C obtained with {\it HST's} Wide Field and Planetary Camera 2 (WFPC2) is consistent with an M7 
dwarf.  Rudy, Rossano, \& Puetter (1996) reported $J$, $H$, and $K$ photometry that suggests an earlier spectral type of M6.  No
spectrum of Gl~105C has yet been published. \\

In this paper, we describe the results from multiple-epoch {\it HST} images of Gl~105AC obtained with WFPC2 and the Near-Infrared
Camera and Multi-Object Spectrometer (NICMOS).  We present photometry for Gl~105C spanning 0.3--$2.3~\mu$m, and we compare the 
WFPC2 and NICMOS colors of Gl~105C to the broadband colors of other late-type M dwarfs.   We report the first 
evidence of Gl~105AC's orbital motion, and we weigh the consistency of this observed motion with the published astrometric orbits
and the latest radial-velocity measurements.  Finally, we discuss the likelihood that Gl~105C alone is responsible for Gl~105A's 
astrometric perturbation and radial-velocity trend.\\

\begin{center}
{\bf 2. OBSERVATIONS AND DATA ANALYSIS} \\
\end{center}

\begin{center}
{\it 2.1 WFPC2 and NICMOS Observations}\\
\end{center}

The WFPC2 images of Gl~105AC reported by GFSU were obtained on UT 1995 January~5 and UT 1995 February~10 as part of a direct
search for faint companions to selected nearby stars (Schroeder \& Golimowski 1996; Schroeder et al.\ 2000).  The pair was 
acquired near the center of the Planetary Camera (PC) and imaged through the F555W (WFPC2 $V$) and F814W (WFPC2 $I$) filters
(Biretta et al.\ 1996).  Subsequent observations with WFPC2 were conducted on UT~1997 December~6 and UT~1998 January~4.  Gl~105AC
was again acquired near the center of the PC, and exposures were recorded through the filters F336W (WFPC2 $U$), F439W (WFPC2 
$B$), F675W (WFPC2 $R$), F850LP ($\lambda_c = 0.91~\mu$m, $\Delta\lambda = 0.10~\mu$m), and F1042M ($\lambda_c = 1.02~\mu$m, 
$\Delta\lambda = 0.04~\mu$m).  Collectively, these images span the traditional {\it UBVRI} sequence plus the Gunn $z$ and 
near-infrared $Z$ bandpasses.  The dates, exposure times, and detector gains for each set of PC images are listed in Table~1.
Unsaturated images of Gl~105A were obtained only in the 0.3~s exposures recorded through F1042M. \\

Near-infrared observations of Gl~105AC were conducted on UT~1998 January~9 as part of an {\it HST} snapshot search for faint
companions to stars within 10~pc of the Sun (Krist et al.\ 1998) using NICMOS Camera~2 (NIC2; Calzetti et al.\ 1999).  These 
NICMOS observations were contemporaneous with our last WFPC2 observations.  Gl~105AC was acquired near the center of the 
NIC2 aperture.  Exposures were recorded in {\small MULTIACCUM} mode through the filters F110W (NICMOS $J$), F180M ($\lambda_c = 
1.80~\mu$m, $\Delta\lambda = 0.07~\mu$m), F207M ($\lambda_c = 2.08~\mu$m, $\Delta\lambda = 0.15~\mu$m), and F222M ($\lambda_c = 
2.22~\mu$m, $\Delta\lambda = 0.14~\mu$m).  Collectively, these bandpasses span the conventional {\it JHK} near-infrared sequence.
The exposure times for each set of NIC2 images are listed in Table~1.  Gl~105A saturated the NIC2 detector in the shortest 
readout time through all four bandpasses. \\

The PC and NIC2 images were flux-calibrated using the Space Telescope Science Data Analysis System (STSDAS) software and the 
calibration reference files recommended by the {\it HST} data archive for each epoch.  The images within each set of PC 
exposures were averaged using a $3\sigma$ rejection algorithm to produce a single image devoid of cosmic-ray artifacts.  The
two F336W images had different exposure times, so they were not combined.  For these images, cosmic-ray artifacts were 
identified by visual inspection. \\

To obtain accurate photometry of Gl~105C, subtraction of the nonuniform background signal from Gl~105A's point-spread function 
(PSF) was necessary.  For the PC images, the local background signal was subtracted using the NOAO IRAF task {\small IMSURFIT}.  
Bivariate Legendre polynomials of varying order were fitted to the PSF outside a circular aperture encompassing Gl~105C and lying
within an $n \times n$ pixel subimage, where $n$ varied from 10 to 50, centered on Gl~105C.  The fitted surfaces were then 
subtracted from the subimage.  For the NIC2 images, the PSFs were subtracted using suitably scaled and registered reference 
images of the K1~{\small V} star Gl~68 ($V = 5.22$) and the K3~{\small V} star Gl~892 ($V = 5.56$).  Details of the selection 
and subtraction of NIC2 reference PSFs are reported by Krist et al.\ (1998). \\

The astrometric and photometric measurements of Gl~105C (all bands) and Gl~105A (F1042M only) were obtained using conventional 
methods of aperture photometry.  The centroids of the PC images were corrected for field distortion using the STSDAS task {\small
METRIC}.  Because the first-epoch images of Gl~105A were saturated, we inferred the star's location by repeatedly marking the 
midline of the diffraction spikes from the secondary-mirror support and then computing the intersection of the orthogonal pairs 
of spikes.  This technique rendered the star's position accurate to $\pm 0.2$ pixel.  The measured PC fluxes were converted to 
Vega-based instrumental magnitudes using the technique of Holtzman et al.\ (1995a) for point source photometry.  The NIC2 fluxes 
were converted to Vega-based instrumental magnitudes using the recipe given in the NICMOS Data Handbook (Dickinson et al.\ 1999).
\\

\begin{center}
{\it 2.2 Radial-Velocity Observations}\\
\end{center}

Radial-velocity measurements of Gl~105A have been obtained over the last 12 years as part of a continuing search for extrasolar
planets conducted at the Lick Observatory 3~m telescope (Marcy \& Butler 1992, 1998).  Radial velocities are determined from
Doppler shifts of the star's echelle spectrum relative to a superimposed reference spectrum of iodine absorption lines with 
accurately known wavelengths.  The reference spectrum is not calibrated against an absolute velocity standard, so the zero point 
of the resulting velocities is arbitrary.  The exposure time for each object spectrum is $\sim 10$ min.  The augmented internal 
Doppler precisions for the measurements made before and after 1994 November are 17~m~s$^{-1}$ and 6.9~m~s$^{-1}$, respectively 
(Cumming et al.\ 1999).\\

\begin{center}
{\bf 3. RESULTS} \\
\end{center}

The PC images of Gl~105AC recorded in 1995 through F555W and F814W were presented by GFSU.  The 1997 and 1998 observations 
through the other WFPC2 filters were conducted with nearly the same {\it HST} roll angle.  Because the later images are similar 
to the previously published images, we do not show them here.  A search of the F1042M and F814W images for other stellar or 
substellar companions within $17''$ of Gl~105A revealed no Gl~105C-like stars and no Gl~229B-like brown dwarfs at separations
greater than $\sim 0$\ad3 and $\sim 3''$, respectively (Schroeder et al.\ 2000).\\

Figure~1 shows the reduced and PSF-subtracted NIC2 images of Gl~105AC recorded through F207M.  Both images are displayed with the
same logarithmic scaling to demonstrate the effectiveness of the PSF subtraction.  Gl~105C appears 3\ad2 to the left ({\it i.e.},
northwest) of the saturated image of Gl~105A.  The residuals from the PSF subtraction in the vicinity of Gl~105C are sufficiently
small to make visible the radial diffraction spikes emanating from Gl~105C's image.  Similar degrees of background subtraction 
were obtained for the images recorded through the other NIC2 filters.  No other point sources appear in the any of the NIC2 
images. \\

For stars brighter than $J \approx 5$, our NICMOS limits for detecting faint companions are set by the quality of the PSF
subtraction everywhere in the field (Krist et al.\ 1998).  The faintest limits are reached in our F110W images.  For Gl~105A
($J \approx 4$), the F110W magnitude detection limits are approximately 15.0, 17.5, and 18.5 for separations of 1\ad5, 3\ad0, and
6\ad0, respectively.  These limits are up to eight magnitudes below the empirical limit of $M_J = 11$ at the low-mass end of the 
main sequence (Henry \& McCarthy 1993).\\

\begin{center}
{\it 3.1 Photometry of Gl 105C}\\
\end{center}

Table~2 lists the WFPC2 and NICMOS instrumental magnitudes and their uncertainties for Gl~105C.  The uncertainties in the apparent
magnitudes represent the following effects combined in quadrature: read noise, photon noise, PSF subtraction error, aperture
correction error, flat-field inaccuracy (1\% on small scales for WFPC2; 3\% for NICMOS), charge-transfer inefficiency ($\sim 2$\%
for WFPC2), and zero-point uncertainty (\la~2\% for WFPC2; 3\% for NICMOS).  The absolute magnitudes and their uncertainties were
computed using a parallax of 0\ad13796~$\pm$~0\ad00090, which is the weighted mean of the values for Gl~105A listed in the latest
releases of the Yale and {\it Hipparcos} catalogues of parallaxes (van Altena, Lee, \& Hoffleit 1995; ESA 1997). \\

The discrepancies between the F555W and F814W magnitudes listed in Table~2 and those reported by GFSU reflect the differences in
zero points (ZPs) between the WFPC2 magnitude system used in this paper and the STScI magnitude (STMAG) system used by GFSU.  
Holtzman et al.\ (1995a) report $\Delta ZP_{(WFPC2-STMAG)} = 0.03$ and $-1.21$ for F555W and F814W, respectively.  These 
zero-point offsets notwithstanding, the magnitudes are the same within the reported uncertainties.  On the other hand, the F675W 
magnitude in Table~2 is $\sim 1.7$ mag brighter than the Cousins $R$ and Gunn $r$ magnitudes reported by GNKO.  The F675W 
measurement is consistent with that of a normal M7 dwarf (see Table~3).  Although Gl~105C may be photometrically variable at 
such wavelengths, we surmise that the ground-based $R$-band measurements reported by GNKO are incorrect. \\

\begin{center}
{\it 3.2 Relative Astrometry of Gl 105C}\\
\end{center}

The positions of Gl~105AC in the PC were measured from the 1~s F814W images and the F1042M images using the techniques described 
in \S2.1.  Both stars were saturated in the 35~s F814W images, so the positions of each on UT~1995~February~10 were not measured.
We also did not measure the positions of the stars in the NIC2 images because these images were contemporaneous with the latest 
PC images and because the NIC2 pixel scale is 65\% larger than the that of the PC.  Adopting an image scale of 0\ad04554 
pixel$^{-1}$ for the PC (Holtzman et al.\ 1995b) and {\it HST} roll angles of 71\dd28 and 75\dd00 for the 1995 and 1997/1998
observations, respectively, we obtain the following separations and position angles for Gl~105AC:\\

\begin{center}
\begin{tabular}{rl}
1995 January 05: & 3\ad394 $\pm$ 0\ad010 at 289\dd65 $\pm$ 0\dd26 \\
1997 December 06: & 3\ad223 $\pm$ 0\ad008 at 293\dd80 $\pm$ 0\dd24 \\
1998 January 04: & 3\ad221 $\pm$ 0\ad008 at 293\dd97 $\pm$ 0\dd24 \\
\end{tabular}
\end{center}

The uncertainties reflect centroid errors of $\sim 0.1$ pixel for unsaturated images, a position error of $\pm~0.2$ pixel for the
saturated F814W image of Gl~105A, and estimated roll-angle errors of $\pm$~0\ad07 determined from the canonical {\it HST}
guide-star position error of $1''$.  The slight difference between the first-epoch astrometry given above and that reported
by GSFU is attributed to our improved techniques for PSF subtraction and computing the centers of saturated images.  {\it Because
Gl~105AC is a well-established common proper motion pair (GNKO, GSFU), we conclude that the relative motion over three years 
tabulated above is orbital.} \\

\begin{center}
{\it 3.3 Radial Velocity Measurements}\\
\end{center}

Figure 2 shows 35 radial-velocity measurements of Gl~105A obtained between epochs 1987.7 and 2000.0.  During this time, the 
velocity of Gl~105A varied almost linearly by +140~m~s$^{-1}$.  A linear least-squares fit to the data yields a slope of
$11.3~\pm~0.8$~m~s$^{-1}$~yr$^{-1}$ with a RMS error of 9.9~m~s$^{-1}$.  The slope and duration of this linear trend
imply that there exists a companion with mass greater than 0.01~\msun. \\

\begin{center}
{\bf 4. DISCUSSION} \\[0.1in]
{\it 4.1 Broadband Spectral Type}\\
\end{center}

The $U$-to-$K$ baseline of our photometry provides good leverage for determining the spectral type of Gl~105C.  Table~3 lists the
optical colors for ten M5.5--M8 dwarfs and Gl~105C.  Despite some systematic differences between the WFPC2 and Johnson--Cousins 
systems, all the colors except $U$--$B$ become redder with increasing spectral subclass.  Our WFPC2 colors indicate that 
Gl~105C's spectral type is M7.  However, Gl~105C's F336W magnitude is brighter by 1.5--2 mag than expected from the $U$ 
magnitudes of stars with spectral types M5.5--M6.  This anomaly is probably caused by a known red leak in the F336W filter, 
which is \ga~40\% for objects with $U$--$B \approx 2$ (Holtzman et al.\ 1995a). \\

Figure~3 shows a color-magnitude diagram for Gl~105C and 12 other late-M dwarfs whose parallaxes are known and for which the same
four-band NICMOS photometry has been obtained.  We selected Johnson $V$ and NICMOS F222M for this diagram because these 
bandpasses mark the longest wavelength baseline over which photometry exists for all 13 stars.  (We used the F555W magnitude 
listed in Table~2 as a close approximation to the Johnson $V$ magnitude of Gl~105C.)  The arrows in the lower left corner of 
Figure~3 represent the boundaries of the empirical scatter in $M_V$ obtained from the best-fit relation of Henry, Kirkpatrick, 
\& Simons (1994) for late M dwarf spectral classes.  Gl~105C's measured brightness ($M_V = 17.5$) lies within the photometric 
uncertainties for types M6.5 and M7, but it is $\sim 2\sigma$ fainter than the $M_V = 16.2$ best-fit value for type M6.  
Moreover, the $(V - F222M)$ color of Gl~105C is over a magnitude redder than those of the M6 dwarfs GJ~1245B and LHS~1375.
Thus, our optical and near-infrared data suggest that the spectral type of Gl~105C is closer to the M7 estimate of GSFU than the
M6 estimate of Rudy et al.\ (1996). \\

\begin{center}
{\it 4.2 Is Gl 105C the Astrometric Companion?}\\
\end{center}

GNKO noted large discrepancies between the observed position of Gl~105C and the positions of the astrometric companion expected 
from the orbital elements of Ianna (1992) and Heintz \& Cantor (1994).  Although the astrometric orbits themselves differ
significantly, the periods of each are similar: 59.5~yr (Ianna 1992) and 61~yr (Heintz \& Cantor 1994).  GSFU noted that, to 
satisfy both the $\sim 60$-yr period and the first-epoch WFPC2 observations, either another unseen companion must exist or 
Gl~105C must be near apastron in a highly eccentric orbit.  The nondetection of other stellar or substellar companions in our
PC and NIC2 images (Schroeder et al.\ 2000; this paper) makes the former possibility unlikely.  Having directly observed the 
orbital motion of Gl~105C, we now investigate the latter possibility. \\

The 3-yr span of our observations is insufficient for computing Gl~105C's orbit, but the basic elements of the orbit can be
constrained from our astrometry and Kepler's laws.  Following the method of Golimowski et al.\ (1998) for Gl~229B, we computed
the ranges of Gl~105C's line-of-sight position and velocity that, together with the star's observed position and velocity in the 
plane of the sky (see \S3.2), satisfy bound Keplerian orbits.  Figure~4 shows the loci of periods ($P$), eccentricities ($e$), 
and semi-major axes ($a$) of these bound orbits, plotted as functions of line-of-sight position and velocity.  Figure~4a reveals 
that $P \approx 60$~yr is possible if, from January 1995 to January 1998, Gl~105C's line-of-sight position and velocity were 
approximately zero.  According to Figures~4b and 4c, such an orbit would have $e \approx 0.75$ and $a \approx 15$~AU.  Given a 
distance to the system of 7.2~pc (see \S3.1), the average projected separation of Gl~105AC over the span of our observations was 
$\sim 24$~AU.  If the preceding values of $P$, $e$, and $a$ are correct, then Gl~105C can indeed be the astrometric companion 
and, during our WFPC2 observations, would have been near apastron in a highly eccentric orbit. \\

These conclusions can be checked against the measurements of Gl~105A's radial velocity over the last 12~yr.  Figure~2 shows that,
during this time, only a linear trend in the radial velocity was detected.  Without perceptible curvature in the velocity data, 
it is not possible to constrain the orbit of Gl~105A.  Nevertheless, the observed velocity trend may be compared with the 
variations expected from a VLM companion in a 60-yr orbit around Gl~105A.  Using a derived mass of 0.81~\msun\ for Gl~105A (Henry
\& McCarthy 1993), we compute that a VLM companion would induce peak-to-peak velocity variations of $\sim 1.75 \times 
10^4~M_{vlm}$~sin~$i$~(m~s$^{-1}$), where $M_{vlm}$ is the mass of the VLM companion (in solar masses) and $i$ is the inclination
of the orbit relative to the plane of the sky.  If we assume that Gl~105A's observed velocity variation of 140~m~s$^{-1}$ over 
12~yr is approximately half its peak-to-peak variation (which is not unreasonable for 20\% coverage of a 60-yr orbit), then 
\begin{equation}
M_{vlm}~{\rm sin}~i \approx 0.016~M_{\odot}.
\end {equation}
\noindent
Using the mass--luminosity relation of Henry et al.\ (1999), we derive a mass of 0.082~\msun\ for Gl~105C.  To satisfy Eq.~(1),
Gl~105C must lie in an orbit with $i \approx 11$\deg, {\it i.e.}, a nearly face-on orbit.  This result is consistent with the 
``approximately zero'' values for line-of-sight position and velocity required for Gl~105C from Figure~4a.  Note, however, that 
Eq.~(1) may be satisfied by a companion of lesser mass in a more inclined orbit.  (Indeed, dropping the $P \approx 60$~yr 
constraint permits an even wider range of possibilities.)  Our observations do not preclude the existence of such a companion, 
but neither do they require it in the presence of Gl~105C. \\

The combined evidence from our WFCP2 observations and radial-velocity measurements supports the notion that Gl~105C is the cause 
of the 60-yr astrometric perturbation of Gl~105A.  Considering also the nondetection of other companions in our PC and NIC2 
images, we find no reason to postulate the existence of a fourth component in the Gl~105 system.  Note that these conclusions 
ignore the computed orbital elements of Ianna (1992) and Heintz \& Cantor (1994) except for $P$.  We accept $P \approx 60$~yr as 
valid because (1) the McCormick and Sproul groups independently derived this value, and (2) of all the computed orbital elements,
$P$ is the least sensitive to uncertainties in the astrometric data. \\

\begin{center}
{\bf 5. SUMMARY AND CONCLUSIONS} \\
\end{center}

We have obtained multicolor images of Gl~105AC over a 3-yr period using WFPC2 and NICMOS.  The optical and near-infrared colors 
of Gl~105C strongly suggest a spectral type of M7~{\small V} for that star.  Relative astrometric measurements reveal the 
first evidence of the Gl~105C's orbital motion.  Previous long-term astrometric studies at Sproul and McCormick Observatories 
have shown that the period of Gl~105A's perturbation is $\sim 60$~yr.  To satisfy both the observed orbital motion and Gl~105A's 
astrometric period, Gl~105C's orbit must have $e \approx 0.75$ and $a \approx 15$~AU.  Measurements of Gl~105A's radial velocity 
over 12~yr show a linear trend with a slope of 11.3~$\pm$~0.8~m~s$^{-1}$~yr$^{-1}$.  This trend is consistent with the orbital 
constraints imposed by our multiple-epoch images and $P \approx 60$~yr.  To account for the observed velocity variations, Gl~105C
must be in a nearly face-on orbit.  As no other faint companions to Gl~105A have been detected, we conclude that Gl~105C is 
probably the source of the 60-yr astrometric perturbation.\\[0.2in]

Support for the WFPC2 work was provided by NASA grants NAG5-1617 and NAG5-1620.  The NICMOS study was funded by NASA through 
grants GO-07420.0$x$-96A ($x = 1$,2,3,4) from the Space Telescope Science Institute, which is operated by the Association of 
Universities for Research in Astronomy, Inc., under NASA contract NAS~5-26555.  Support for the Lick Observatory radial-velocity
program is provided by NASA grant NAG5-8299 and NSF grant AST95-20443 (to GWM).

\pagebreak
\begin{center}

{\bf REFERENCES}
\end{center}
\noindent
Biretta, J.~A.\ et al.\ 1996, WFPC2 Instrument Handbook, Version 4.0 (Baltimore: STScI) \\
Calzetti, D.\ et al.\ 1999, NICMOS Instrument Handbook, Version 3.0 (Baltimore: STScI) \\
Cumming, A., Marcy, G.~W., \& Butler, R.~P. 1999, ApJ, 526, 890 \\
Dickinson, M.\ et al.\ 1999, NICMOS Data Handbook, Version 4.0 (Baltimore: STScI) \\
ESA 1997, The {\it Hipparcos} and Tycho Catalogues, ESA SP-1200 (Noordwijk: ESA) \\
Golimowski, D.~A., Burrows, C.~J., Kulkarni, S.~R., Oppenheimer, B.~R., \& Brukardt, R.~A. \linebreak\hspace*{0.5in} 1998, AJ, 
115, 2579 \\
Golimowski, D.~A., Fastie, W.~G., Schroeder, D.~J., \& Uomoto, A. 1995a, ApJ, 452, L125 \linebreak\hspace*{0.5in}(GFSU)\\
Golimowski, D.~A., Nakajima, T., Kulkarni, S.~R., \& Oppenheimer, B.~R. 1995b, ApJ, \linebreak\hspace*{0.5in}444, L101 
(GNKO)\\
Hawley, S.~L., Gizis, J.~E., \& Reid, I.~N. 1996, AJ, 112, 2799 \\
Hawley, S.~L., Gizis, J.~E., \& Reid, I.~N. 1997, AJ, 113, 1458 \\
Heintz, W.~D., \& Cantor, B.~A. 1994, PASP, 106, 363 \\
Henry, T.~J., Franz, O.~G., Wasserman, L.~H., Benedict, G.~F., Shelus, P.~J., Ianna, P.~A., \linebreak\hspace*{0.5in}
Kirkpatrick, J.~D., \& McCarthy, D.~W. 1999, ApJ, 512, 864 \\
Henry, T.~J., Kirkpatrick, J.~D., \& Simons, D.~A. 1994, AJ, 108, 1437 \\
Henry, T. J., \& McCarthy, D. W. 1993, AJ, 106, 773\\
Holtzman, J.~A., Burrows, C.~J., Casertano, S., Hester, J.~J., Trauger, J.~T., Watson, A.~M., \linebreak\hspace*{0.5in} \& 
Worthey, G. 1995a, PASP, 107, 1065 \\
Holtzman, J., Hester, J.~J., Casertano, S., Trauger, J.~T., Watson, A.~M., Ballester, G.~E., \linebreak\hspace*{0.5in} Burrows, 
C.~J., Clarke, J.~T., Crisp, D., Evans, R.~W., Gallagher, J.~S., Griffiths, \linebreak\hspace*{0.5in} R.~E., Hoessel, J.~G., 
Matthews, L.~D., Mould, J.~R., Scowen, P.~A., Stapelfeldt, \linebreak\hspace*{0.5in} K.~R., \& Westphal, J.~A. 1995b, PASP, 107, 
156 \\
Ianna, P.~A. 1992, in ASP Conf.\ Ser.\ 32, Complementary Approaches to Double and Multiple \linebreak\hspace*{0.5in} Star 
Research, ed.\ H.~A.\ McAlister \& W.~I.\ Hartkopf (San Francisco: ASP), 323 \\
Kirkpatrick, J.~D., Henry, T.~J., \& Simons, D.~A. 1995, AJ, 109, 797 \\
Krist, J.~E., Golimowski, D.~A., Schroeder, D.~J., \& Henry, T.~J. 1998, PASP, 110, 1046 \\
Leggett, S.~K. 1992, ApJS, 82, 351 \\
Lippincott, S.~L. 1973, AJ, 78, 303 \\
Marcy, G.~W., \& Butler, R.~P. 1992, PASP, 104, 270 \\
Marcy, G.~W., \& Butler, R.~P. 1998, ARA\&A, 36, 57 \\
Martin, G.~E., \& Ianna, P.~A. 1975, AJ, 80, 321 \\
Rudy, R.~J., Rossano, G.~S., \& Puetter, R.~C. 1996, ApJ, 458, L41 \\
Schroeder, D.~J., \& Golimowski, D.~A. 1996, PASP, 108, 510 \\
Schroeder, D.~J, Golimowski, D.~A., Brukardt, R.~A., Burrows, C.~J., Caldwell, J.~J., Fastie, \linebreak\hspace*{0.5in}W.~G., 
Ford, H.~C., Hesman, B., Kletskin, I., Krist, J.~E., Royle, P., \& Zubrowski, \linebreak\hspace*{0.5in}R.~A. 2000, AJ, 119, 906. 
\\
van Altena, W.~F., Lee, J.~T., \& Hoffleit, E.~D. 1995, The General Catalogue of Trigono- \linebreak\hspace*{0.5in} metric 
Stellar Parallaxes, Vols.\ 1 and 2, Fourth Edition (New Haven: Yale Univ.\ \hspace*{0.5in} Obs.) \\
van Maanen, A. 1938, ApJ, 88, 277 \\
Weis, E.~W. 1996, AJ, 112, 2300 \\

\newpage 
{\small
\begin{center}
TABLE 1\\
WFPC2 and NICMOS Exposures of Gl 105AC\\[0.1in]
\begin{tabular}{ccccrccl}
\hline\hline
       &             &          &         & Exposure & No. of    & Gain \\
Epoch  & UT          & Camera	& Filter  & Time (s) & Exposures & ($e^-$~DN$^{-1}$) & Comments\\
\hline
1      & 1995 Jan 05 & PC	& F555W   & 35~~     & 4         & 14                & WFPC2 $V$ \\
       &             &          & F814W   & 1~~      & 4         & 14                & WFPC2 $I$ \\
       &             &          & F814W   & 35~~     & 8         & 14                & Gl 105C saturated\\
       & 1995 Feb 10 & PC       & F814W   & 35~~     & 8         & 14                & Gl 105C saturated\\
& \\
2      & 1997 Dec 06 & PC       & F675W   & 12~~     & 3         & 7                 & WFPC2 $R$ \\
       &             &          & F850LP  & 4~~      & 3         & 7                 & $\sim$ Gunn $z$ \\
       &             &          & F1042M  & 0.3      & 3         & 14                & Gl 105A unsaturated \\
       &             &          & F1042M  & 260~~    & 3         & 7                 & $\sim$ $Z$\\
       & 1998 Jan 04 & PC       & F336W   & 100~~    & 1         & 7                 & WFPC2 $U$ \\
       &             &          & F336W   & 160~~    & 1         & 7                 & WFPC2 $U$ \\
       &             &          & F439W   & 100~~    & 2         & 7                 & WFPC2 $B$ \\
       &             &          & F1042M  & 0.3      & 3         & 14                & Gl 105A unsaturated \\
       &             &          & F1042M  & 260~~    & 3         & 7                 & $\sim$ $Z$\\
       & 1998 Jan 09 & NIC2     & F110W   & 64~~     & 2         & 5.4               & NICMOS $J$ \\
       &             &          & F180M   & 64~~     & 2         & 5.4               & Narrow $H$ \\
       &             &          & F207M   & 128~~    & 2         & 5.4               & Blue-half of $K$ \\
       &             &          & F222M   & 128~~    & 2         & 5.4               & Red-half of $K$ \\
\hline\hline
\end{tabular}
\end{center}
}
       
\newpage 
\begin{center}
TABLE 2\\
WFPC2 and NICMOS Magnitudes$^a$ of Gl 105C\\[0.1in]
\begin{tabular}{cccc}
\hline\hline
       &        & Apparent         & Absolute \\
Camera & Filter & Magnitude        & Magnitude$^b$ \\
\hline
PC     & F336W$^c$  & $18.99 \pm 0.06$ & $19.69 \pm 0.07$\\
       & F439W  & $19.17 \pm 0.05$ & $19.87 \pm 0.06$ \\
       & F555W  & $16.77 \pm 0.08$ & $17.47 \pm 0.10$ \\
       & F675W  & $14.68 \pm 0.08$ & $15.38 \pm 0.10$ \\
       & F814W  & $12.26 \pm 0.03$ & $12.96 \pm 0.05$ \\
       & F850LP & $11.17 \pm 0.03$ & $11.87 \pm 0.05$ \\
       & F1042M & $10.49 \pm 0.03$ & $11.19 \pm 0.05$ \\
       & \\
NIC2   & F110W  & $10.07 \pm 0.05$ & $10.77 \pm 0.06$\\
       & F180M  & ~$9.18 \pm 0.05$ & ~$9.88 \pm 0.06$ \\
       & F207M  & ~$8.96 \pm 0.05$ & ~$9.66 \pm 0.06$ \\
       & F222M  & ~$8.65 \pm 0.05$ & ~$9.35 \pm 0.07$ \\
\hline\hline
\end{tabular}
\end{center}
\hspace*{1.25in} {\small $^a$ Vega is defined to have zero magnitude in each bandpass.}\\
\hspace*{1.25in} {\small $^b$ Computed using a parallax of 0\ad13796~$\pm$~0\ad00090 (see \S3.1).}\\
\hspace*{1.25in} {\small $^c$ Magnitudes affected by red leak (Holtzman et al.\ 1995a).}
       
\newpage 
\begin{center}
TABLE 3\\
Optical Colors$^a$ of Late M Dwarfs\\[0.1in]
\begin{tabular}{lccccccc}
\hline\hline
        & Spectral \\
Name    & Type            & $U$--$B$ & $B$--$V$ & $V$--$R$ & $R$--$I$ & $V$--$I$ & References$^b$\\
\hline
GJ 1002 & M5.5~{\small V} & 1.88     & 1.97     & 1.59     & 2.01     & 3.60     & 1,2,3,3,3,3\\
Gl 551  & M5.5~{\small V} & 1.37     & 1.90     & 1.65     & 2.00     & 3.65     & 4,2,2,2,2,2\\
Gl 905  & M5.5~{\small V} & 1.45     & 1.91     & 1.52     & 1.95     & 3.47     & 1,2,3,3,3,3\\
Gl 406  & M6.0~{\small V} & 1.59     & 2.00     & 1.87     & 2.19     & 4.06     & 1,2,3,3,3,3\\
GJ 1245B& M6.0~{\small V} & --       & 1.97     & 1.65     & 2.09     & 3.74     & 1,-,2,3,3,3\\
LHS 292 & M6.5~{\small V} & --       & 2.10     & 2.20     & 2.22     & 4.42     & 1,-,2,3,3,3\\
GJ 1111 & M6.5~{\small V} & --       & 2.06     & 2.01     & 2.23     & 4.24     & 1,-,2,3,3,3\\
Gl 644C & M7.0~{\small V} & --       & 2.20     & 2.15     & 2.41     & 4.56     & 1,-,2,2,2,2\\
LHS 3003& M7.0~{\small V} & --       & --       & 2.17     & 2.35     & 4.52     & 5,-,-,2,2,2\\
Gl 752B & M8.0~{\small V} & --       & 2.13     & --       & --       & 4.70     & 1,-,2,2,-,2\\
\hline
Gl 105C & M7.0~{\small V} & $-0.18^c$  & 2.40     & 2.09     & 2.42     & 4.51     & 6,6,6,6,6,6\\
\hline\hline
\end{tabular}
\end{center}
{\small
\hspace*{0.50in} $^a$ WFPC2 colors given for Gl~105C; Johnson--Cousins colors given for all others.\\
\hspace*{0.50in} $^b$ Spectral type, $U$, $B$, $V$, $R$, $I$.\\
\hspace*{0.50in} $^c$ Color affected by red leak in F336W filter (Holtzman et al.\ 1995a).\\
\hspace*{0.50in} References: (1) Henry, Kirkpatrick, \& Simons 1994; (2) Leggett 1992; (3) Weis 1996; \\
\hspace*{1.30in} (4) Hawley, Gizis, \& Reid 1996, 1997; (5) Kirkpatrick, Henry, \& Simons \\
\hspace*{1.30in} 1995; (6) this paper.\\
}
       
\newpage
\begin{center}
{\bf Figure Captions}\\
\end{center}

\noindent
{\bf Figure 1.} NICMOS Camera 2 (NIC2) image of Gl~105AC recorded through F207M.  The logarithm of the image is shown to reduce 
image contrast.  The panels depict the calibrated image before {\it (left)} and after {\it (right)} subtraction of the primary 
star's PSF using a reference image of Gl~892 (K3~{\small V}, $V = 5.56$).  Gl~105C lies 3\ad2 to the left ({\it i.e.}, northwest)
of Gl~105A.  No other point sources appear in the 19\ad2~$\times$~19\ad2 NIC2 field of view. \\

\noindent
{\bf Figure 2.} Measured radial velocities of Gl~105A over the past 12 years.  The data follow a linear trend with slope 
$11.3~\pm~0.8$~m~s$^{-1}$~yr$^{-1}$. \\

\noindent
{\bf Figure 3.} Color--magnitude diagram for Gl~105C and 12 dwarfs with known parallaxes and spectral types M5 {\it (circles)},
M5.5 {\it (squares)}, M6 {\it (triangles)}, M6.5 {\it (diamond)} and $>$M7 {\it (inverted triangle)}.  Johnson $V$ and NICMOS 
F222M magnitudes are used for all stars except Gl~105C.  The F555W magnitude from Table~2 is used as a close approximation for 
Gl~105C's $V$ magnitude.  The arrows at left represent the boundaries of the empirical scatter in $M_V$ obtained from the 
best-fit relation of Henry et al.\ (1994) for dwarf spectral types M6 {\it (short-dashed line)}, M6.5 {\it (long-dashed line)}, 
and M7 {\it (solid line)}.  The 12 dwarfs represented are (1)~LHS~3262, (2)~GJ~1253, (3)~GJ~1057, (4)~GJ~1154, (5)~LHS~1809, 
(6)~GJ~1245A, (7)~GJ~1286, (8)~GJ~1245B, (9)~LHS~1326, (10)~LHS~1375, (11)~LHS~2930, and (12)~GJ~1245C.  The photometric 
uncertainties for these stars are comparable to those shown for Gl~105C.\\

\noindent
{\bf Figure 4.}  Loci of {\it (a)} periods, {\it (b)} eccentricities, and {\it (c)} semimajor axes consistent with bound 
Keplerian orbits and the observed motion of Gl~105C between January 1995 and January 1998.  The parameters are shown as functions
of the line-of-sight position and velocity of Gl~105C relative to the plane of the sky.  The outer contours reflect the 
boundaries between bound and unbound orbits.  The astrometric period of 60~yr is satisfied if Gl~105C's orbit has $e \approx 
0.75$ and $a \approx 15$~AU.

\end{typestyle}
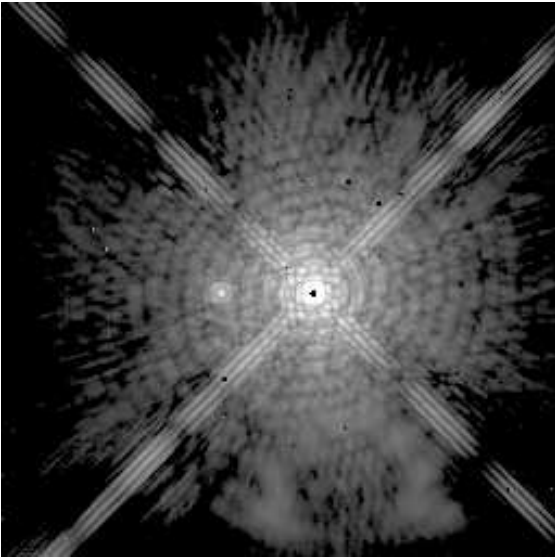
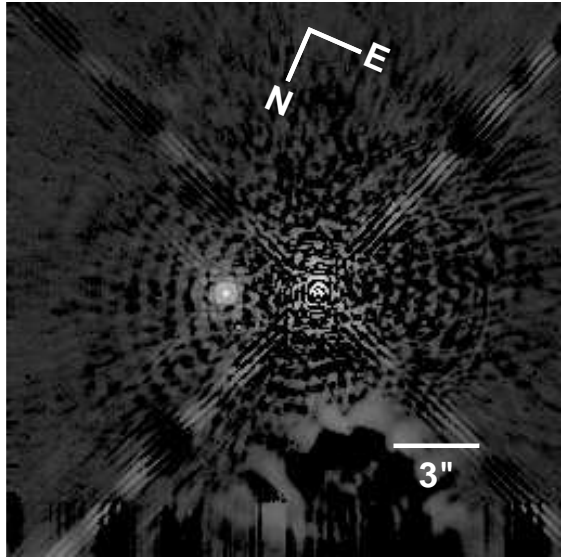
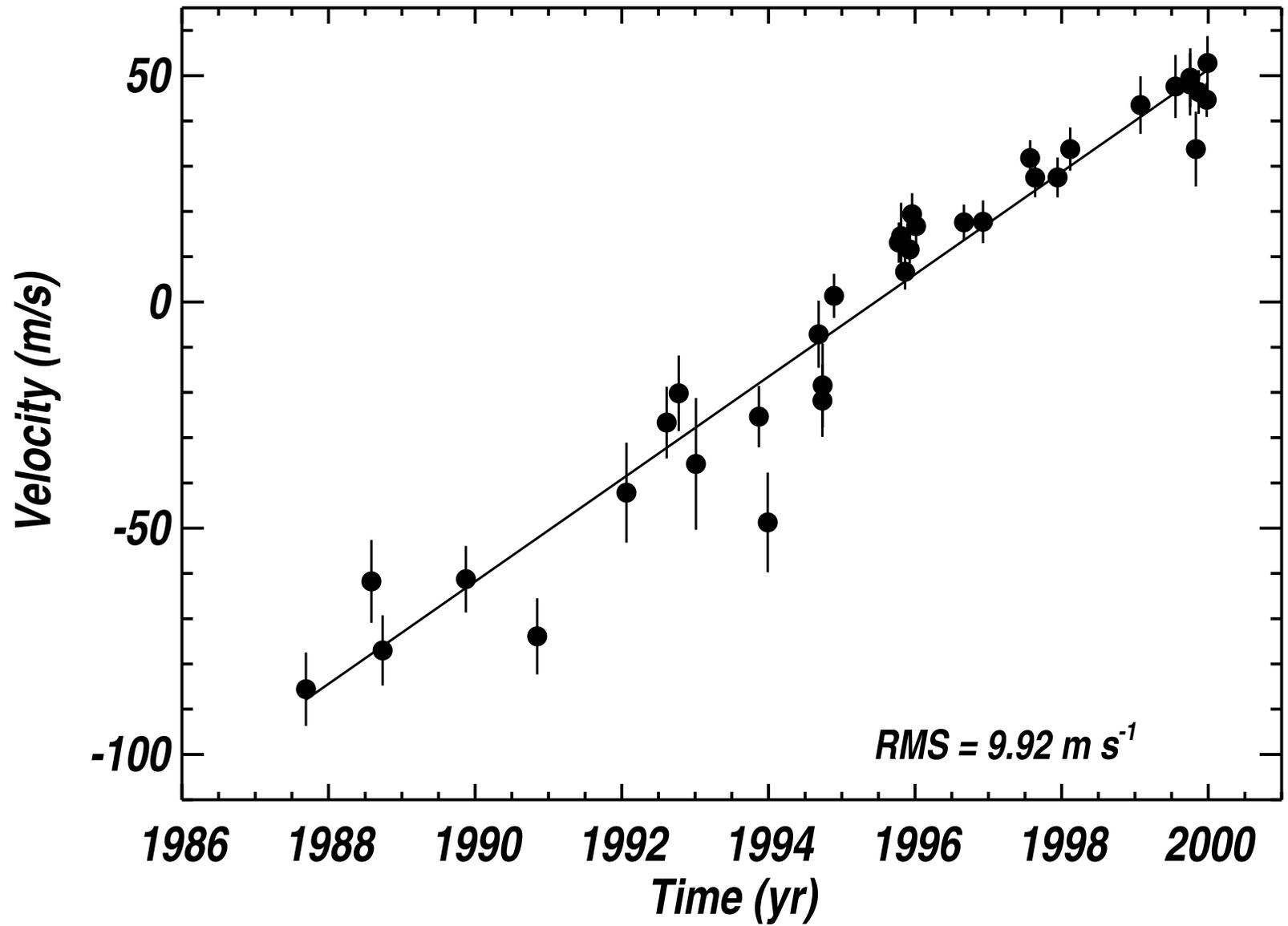
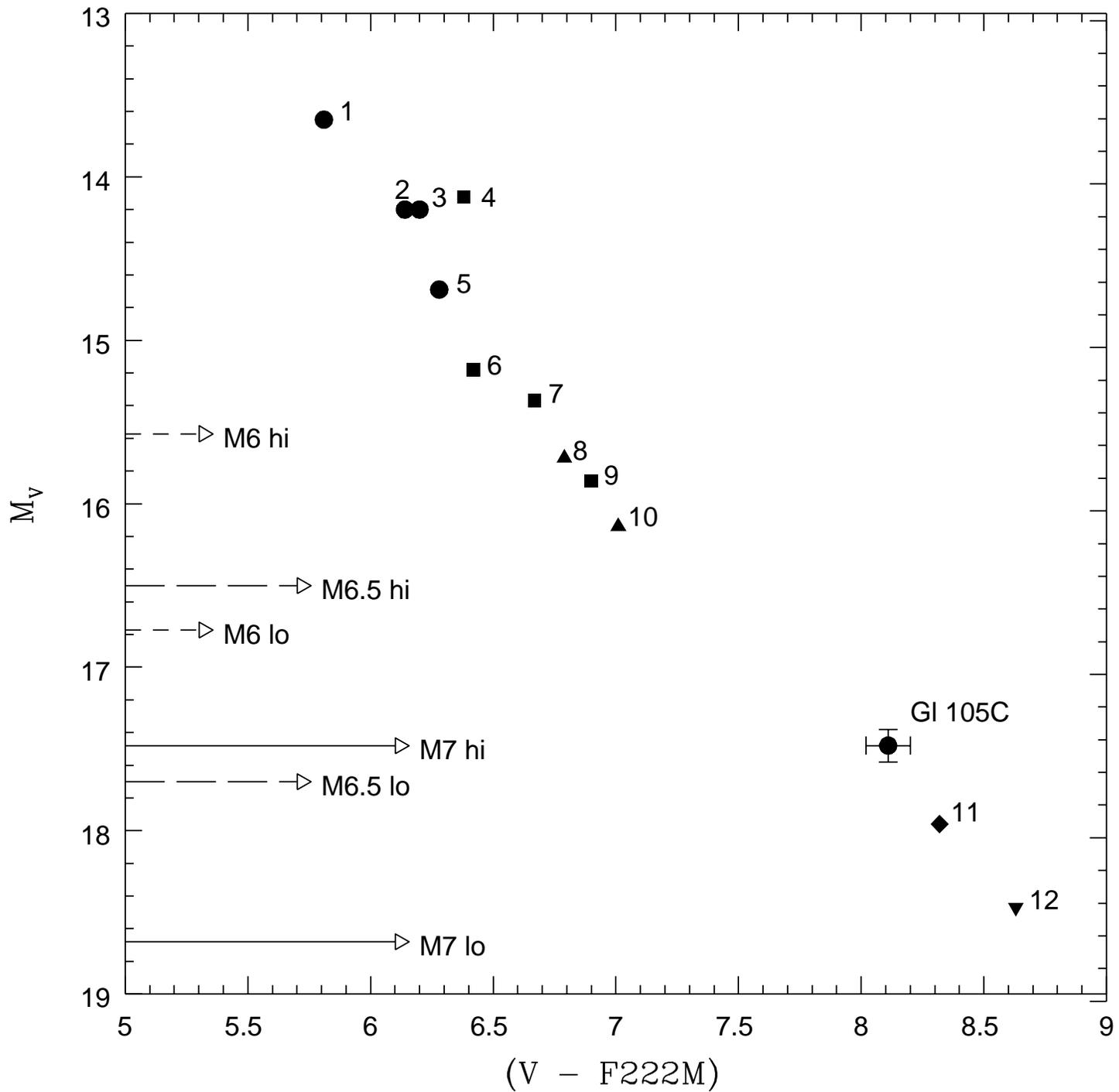
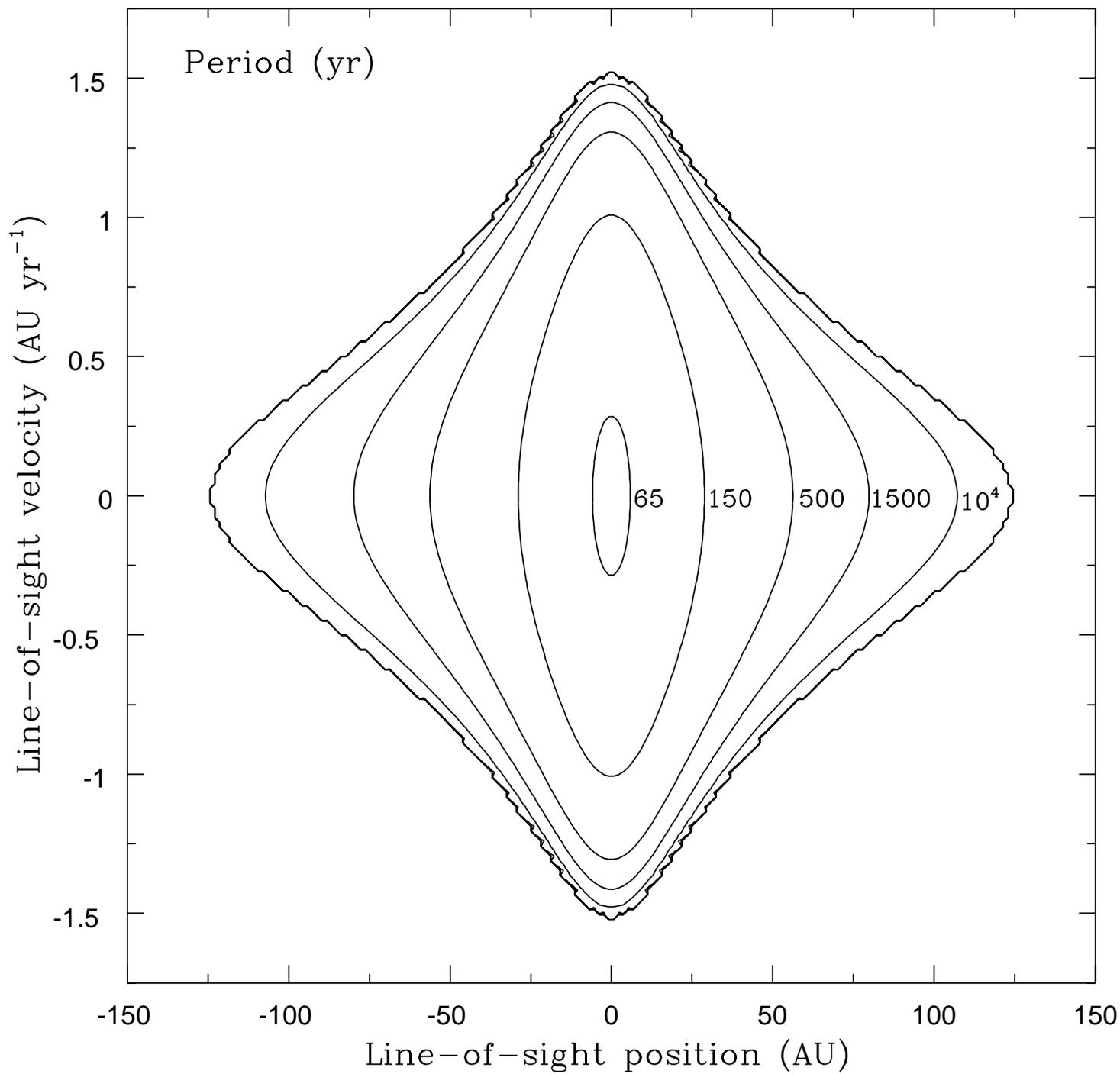
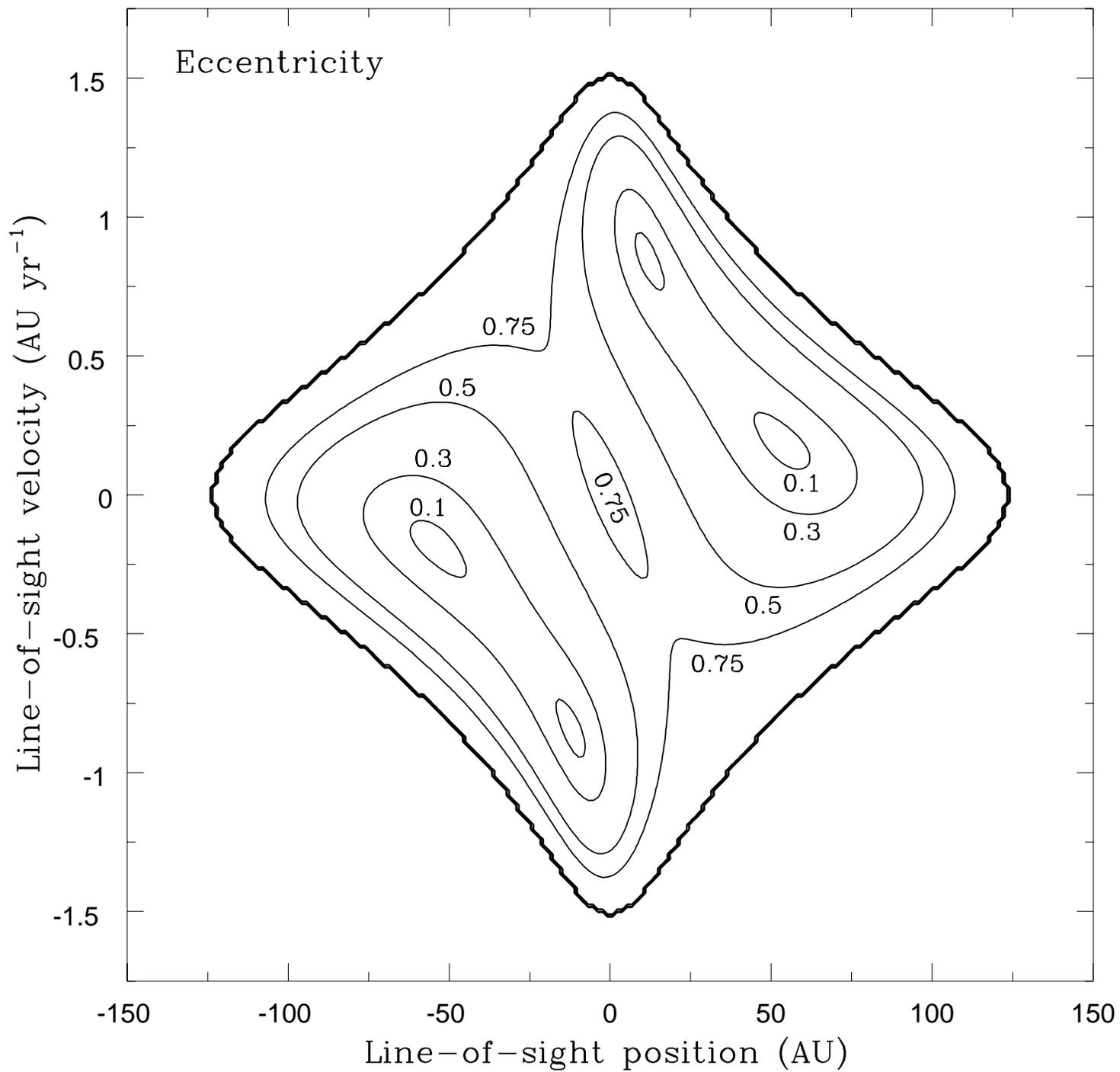
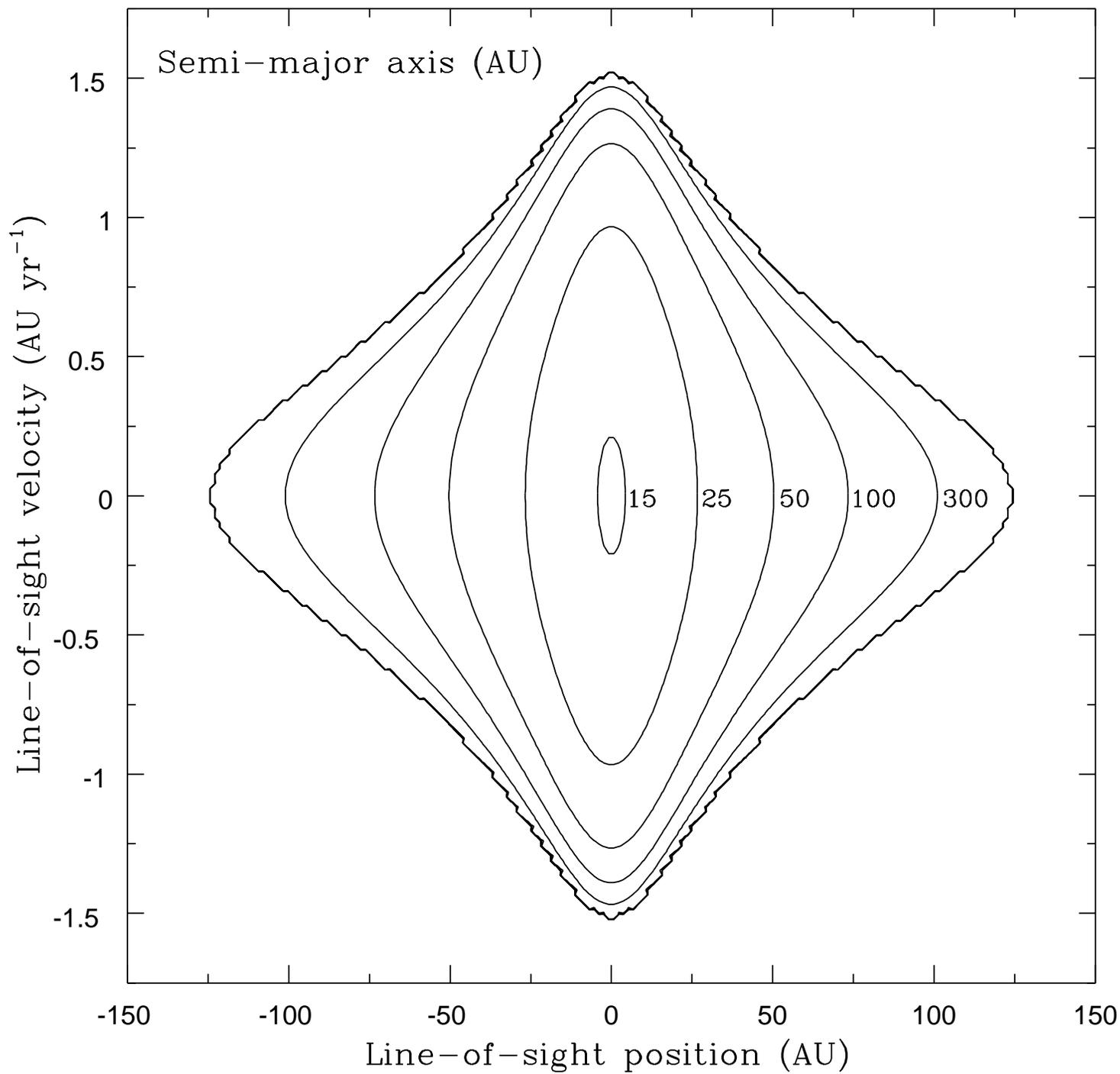
\end{document}